\documentstyle[preprint,aps,epsfig,axodraw]{revtex}

\def\PLB#1#2#3{{\em Phys. Lett.} {\bf B#1} (19#2) #3}

\def\PRL#1#2#3{{\em Phys. Rev. Lett.} {\bf#1} (19#2) #3}

\begin{document}
\newcommand{\cross}{\mbox{$\rlap{\kern0.125em/}$}}
\draft
\preprint{\vbox{\hbox{USM-TH-124}}}
\title{Classical sum rules and spin correlations in photoabsorption
and photoproduction processes}
\author{Iv\'an Schmidt, Alfonso R. Zerwekh }
\address{
 {\it Departamento de F{\'{\i}}sica,
Universidad T\'ecnica Federico Santa Mar{\'{\i}}a, \\ Casilla
110-V, Valpara\'\i so, Chile} }
\maketitle
\widetext
 \begin{abstract}
   In this paper we study the possibility of  generalizing the
classical photoabsorption  ($\gamma a \to b c$) sum rules, to
processes  $b c \rightarrow \gamma a$ and crossed helicity
amplitudes. In the first case, using detailed balance, the  sum
rule is written as $\int_{\nu _{th} }^\infty {\frac{{d\nu }}{\nu
}} K\Delta \sigma _{Born} (\nu )=0$ where $K$ is a kinematical
constant which depends only on the mass of the particles and the
center of mass energy. For other crossed helicity amplitudes, we
show that there is a range of values of $s$ and $t$ for which the
differential cross section for the process $\gamma b \to a c$ or
$a c \to \gamma b$ in which the helicities of the photon and
particle $a$ have specific values, is equal to the differential
cross section for the process in which one of these two helicities
is reversed (parallel-antiparallel spin correlation).

\end{abstract}

\section{Introduction}
\label{sec:int}

Deep inelastic and photon scattering sum rules have been
historically an important tool for obtaining information about
hadron structure \cite{LlewellynSmith:1998wc}. Some of them, such
as the Bjorken polarized deep inelastic sum rule, are exact QCD
results, and therefore constitute stringent tests of this theory.
Others are based on various assumptions which in some cases have
been proved to be incorrect, but even here these sum rules have
been useful. For example, the failure of the Gottfried sum rule
gave essential information about the proton sea. More recently,
the failure of the Ellis-Jaffe sum rule indicated the presence of
polarized strange quarks inside the nucleon.

In 1965 Drell and Hearn \cite{dh}, and independently Gerasimov
\cite{gerasimov}, found an amazing equation that relates the
square of the anomalous magnetic moment of a particle to the
logarithmic integral of the difference of the cross section for
the absorption of a photon with spin parallel or anti-parallel to
the target spin:

\begin{equation}
\mu_a^2={4\pi\alpha S \over M^2} (g-2)^2= {S \over \pi}
\int _{\nu _{th}}^{\infty }{d\nu \over
\nu }\ \Delta\sigma (\nu ),
\label{E1}
\end{equation}
for a target of spin $S$, whether elementary or composite.

This sum rule has interesting theoretical implications. For
example, in pure QED $\Delta\sigma$ is of order $\alpha^2$
(elastic Compton scattering), which means that $g-2 = 0 +
\cal{O}(\alpha)$, and therefore $g = 2$ at tree level for a
particle of any spin. But then:
\begin{equation}
\int _{\nu _{th}}^{\infty }{d\nu \over \nu }\ \Delta\sigma
_{QED}(\nu )\ =\ \cal{O}(\alpha^{\rm 3}), \label{E2}
\end{equation}

This remarkable result was derived by Altarelli, Cabibbo and
Maiani  \cite{altarelli}, and later generalized using loop
counting arguments \cite{bs} to any process of the form $\gamma a
\rightarrow b c$:
\begin{equation}
\label{cdhg}
\int _{\nu _{th}}^{\infty }{d\nu \over \nu }\ \Delta\sigma^{tree}
_{\gamma a \to b c}(\nu )\ = \int\limits_{\nu _{th} }^\infty
\frac{{d\nu }}{\nu } ( \sigma _{P}^{\gamma a \rightarrow b
c}(\nu)-\sigma _{A}^{\gamma a \rightarrow b c}(\nu))=\ 0,
\label{E3}
\end{equation}
where the initial photon and particle $a$ are polarized (parallel
(P) or antiparallel (A) to each other). This result provides
non-trivial checks on calculations, and is a potential diagnostic
tool for new physics. A further consequence is the fact that the
vanishing of this logarithmic integral also implies that there
must be a center of mass energy where $\Delta\sigma(s)$ vanishes,
a result which can be used in phenomenological studies. In fact,
it has been applied to the process $\gamma e \to W \nu$ to probe
anomalous trilinear gauge couplings $\gamma WW$
\cite{Brodsky:1995ga}, in heavy quark electro and photoproduction
reactions \cite{Brodsky:1997ex,Bass:1998rn}, in studies of the
spin-dependent photon structure function \cite{Bass:1998bw}, and
even in quantum gravity \cite{Goldberg:1999gc}.

In this paper, we study the possibility of generalizing the later
results for processes of the form $b c \rightarrow \gamma a$, $a c
\rightarrow \gamma b$ or $\gamma c \rightarrow a b$. In all cases
we consider the correlation between the helicities of the photon
and particle ``$a$''.

The paper is organized as follow.In section II we obtain a
DHG-like sum rule for the process $b c \rightarrow \gamma a$.
Section III is devoted to the study of some properties of crossed
helicity amplitude in order to determine whether it is possible to
build a DHG-like sum rule for processes of the form $a c
\rightarrow \gamma b$ or $\gamma c \rightarrow a b$. We finish
summarizing our main conclusions.

\section{A classical photoproduction sum rule}
\label{sec:1}

\subsection{Deduction of the Sum Rule}
\label{sec:deduction}


The processes $\gamma a \rightarrow b c$ and $b c \rightarrow
\gamma a$ are described by the same amplitude and their center of
mass systems are the same, so in general we can write, using
detailed balance:

\begin{equation}
\label{sigma}
\sigma(\gamma a \rightarrow b c)  =K \sigma(b c \rightarrow \gamma a)
\end{equation}
where $K$ is a constant that depends only on the masses and the
center of mass energy and which takes into account the different
phase spaces of the processes. This constant can be easily
calculated in the center of mass system and we obtain:

\begin{equation}
K = \frac{1}{{\left( {1 - \frac{{m_a^2 }}{s}} \right)^2 }}\left(
{1 - 2\frac{{\left( {m_b^2  + m_c^2 } \right)}}{s} + \frac{{\left(
{m_b^2  - m_c^2 } \right)^2 }}{{s^2 }}} \right).
\end{equation}

Therefore equation (\ref{cdhg}) becomes:

\begin{equation}
\label{result}
\int\limits_{\nu _{th} }^\infty  \frac{{d\nu }}{\nu }K
 ( \sigma _{P}^{b c \rightarrow \gamma a}(\nu)-\sigma _{A}^{b c
 \rightarrow \gamma a}(\nu))=0.
\end{equation}
In this case $\sigma _{P}$  and $\sigma _{A}$ refer to the cross
sections for parallel and anti-parallel spins for the final state
photon and particle $a$, and $\nu$ is the energy of the photon.

Because the kinematical constant $K$ is defined as a ratio of
cross sections (see eq. (\ref{sigma})), it cannot be zero while
the two channels are open. In fact $K$ is always positive. This
observation and equation (\ref{result}) imply that the quantity $(
\sigma _{P}^{b c \rightarrow \gamma a}(\nu)-\sigma _{A}^{b c
\rightarrow \gamma a}(\nu))$ must vanish at a certain energy
$\nu=\nu_{c}$.

A further consequence is that in the photoabsorption process
$\gamma a \to b c$ there is a range of values of the Mandelstam
variables $s=s_0$ and $t=t_0$ for which the differential cross
sections for the processes where the initial photon and particle
$a$ are polarized parallel and antiparallel to each other are
equal:

\begin{equation}
\label{dcs} \left({d\sigma \over dt}\right)_P (s_0,t_0) =
\left({d\sigma \over dt}\right)_A (s_0,t_0),
\end{equation}
and the same holds for the photoproduction process $b c \to \gamma
a$. A specific example of these results will be shown in the next
section.

Thus we have obtained a new sum rule for final state polarized
photons, which implies a zero in the corresponding difference of
total and differential cross sections.
\subsection{An Illustrative Example}
\label{sec:ex}

In order to illustrate the previous results, we consider the
electron-positron annihilation to a pair of photons. In this case
the difference between the cross sections for parallel and
anti-parallel spins in the final states in the center of mass
system can be written as:

\begin{equation}
\label{ex1} \sigma _P  - \sigma _A  = \frac{{e^4 }}{{32\pi m^2
}}\frac{{(1 - y^2 )}}{{y^2 }}\left[ {3y - \log \left(\frac{{1 +
y}}{{1 - y}}\right)} \right]
\end{equation}
where $m$ is the mass of the electron and $y$ is defined as:

\begin{equation}
\label{ex2}
y = \sqrt {1 - \frac{{4m^2 }}{s}}
\end{equation}

In this example the kinematical constant $K$ and the photon energy
$\nu$ take the simple form:

\begin{equation}
  K=y^2
\end{equation}

\begin{equation}
\nu  = \frac{m}{{\sqrt {1 - y^2 } }}
\end{equation}

With these ingredients we can write:

\begin{equation}
\int\limits_0^\infty  {\frac{{d\nu }}{\nu }} K\left( {\sigma _P  -
    \sigma _A } \right) = \frac{{e^4 }}{{32\pi m^2 }}\int\limits_0^1
{dyy\left[ {3y - \log \left( {\frac{{1 + y}}{{1 - y}}} \right)}
  \right]}  = 0
\end{equation}

 From eq. (\ref{ex1}) we can see the the the quantity $\sigma _P  -
\sigma _A$ vanishes for $y \approx 0.859$ or, using (\ref{ex2}),
$\sqrt{s} \approx 7.63 m$.

\section{Helicity Correlation Between an Initial (Final) Polarized Photon
and a Final (Initial) Particle}
 \label{sec:2}

Let us consider now processes of the form $b c \rightarrow \gamma
a$ and $\bar{a} c \rightarrow \gamma \bar{b}$ (crossed), or
equivalently using detailed balance $\gamma a \rightarrow b c$ and
$\gamma \bar{b} \rightarrow \bar{a} c$. It is evident that
equation (\ref{result}) implies the existence of some $s=s_0$ and
$t=t_0$ which satisfy

\begin{equation}
\label{deltam}
\left| {{\cal M}^{b c \to \gamma a} } \right|_P^2 \left( {s_0 ,t_0 }
\right) - \left| {{\cal M}^{b c \to \gamma a} } \right|_A^2
\left( {s_0 ,t_0 } \right) = 0
\end{equation}
\noindent where ${\cal M}^{b c \to \gamma a}$ designates the
amplitude of the process $b c \rightarrow \gamma a$. As noted
before, this means that the corresponding differential cross
sections are also equal at the same values $s_0$ and $t_0$. In
order to study the crossed processes it is necessary to know how
the left part of the previous equation transforms under crossing.

It is a known result that helicity amplitudes transform under crossing
as \cite{martin}:

\begin{equation}
{\cal M}_{bc;\gamma a} (s,t) = \sum\limits_{a',b',c',\gamma '}
{d_{\gamma '\gamma }^{s_\gamma  } (\chi _\gamma  )} d_{a'a}^{s_a }
(\chi _a )d_{b'b}^{s_b } (\chi _b )d_{c'c}^{s_c } (\chi _c )\tilde
{\cal M}_{ca;\gamma b} (s,t)
\end{equation}
\noindent
where $\gamma,a,b,c$ are helicity indexes of the
particles represented by the same letter, $s_i$ represents the
spin of the particle ``$i$'' , $d(\chi)$ are the Wigner matrices
and ${\cal M}$ ($\tilde{\cal M}$) represents the amplitude of the
original (crossed) process.

 From the previous equation it is easy to prove that the left part of
equation (\ref{deltam}) transforms as:

\begin{equation}
  \label{eq:transdelm}
  \left| {{\cal M} } \right|_P^2 \ - \left| {{\cal M} } \right|_A^2 =
\kappa \left( \left|{\tilde {\cal M} } \right|_P^2 \ - \left| {\tilde {\cal M} }
\right|_A^2 \right)
\end{equation}
\noindent
where

\[
\kappa  = \left\{ \begin{array}{cl}
  - 1&{\mbox{ if $a$ is a fermion}} \\
 1&{\mbox{ if $a$ is a massless vector}} \\
 \cos (\chi _a )&{\mbox{ if $a$ is a massive vector}} \\
 \end{array} \right.
\]
\noindent
with, in our case,

\begin{equation}
  \cos (\chi _a )=
\frac{{(s + m_a^2 )(t + m_a^2  - m_b^2 ) - 2m_a^2 (m_a^2  - m_b^2
+ m_c^2 )}}{{(s - m_a^2 )\sqrt {[t - (m_a  + m_b )^2 ][t - (m_a  - m_b )^2 ]} }}
\end{equation}

As an example of the meaning of equation (\ref{eq:transdelm}) ,
consider the processes $e^+ e^- \rightarrow \gamma \gamma$ and $
e^- \gamma \rightarrow e^- \gamma$, where we will study the
 helicity correlation between the photons. It can be shown that:

\begin{eqnarray}
\label{ej1}
 && \left| {{\cal M} } \right|_P^2 \ - \left| {{\cal M} } \right|_A^2
  = \nonumber \\
 && \frac{{2e^4 }}{{(t - m^2 )^2 (u - m^2 )^2 }}\left\{ {(tu - m^4
)(6m^4 + t^2  + u^2  - 4m^2 t + 4m^2 u)
- m^2 s^2 (s - 2m^2)}\right\}
\end{eqnarray}

\begin{eqnarray}
\label{ej2}
 && \left| {\tilde {\cal M} } \right|_P^2 \ - \left| {\tilde {\cal M} } \right|_A^2
  = \nonumber \\
&&\frac{{2e^4 }}{{(\tilde s - m^2 )^2 (\tilde u - m^2 )^2 }}\left\{
  {(\tilde s \tilde u - m^4 )(6m^4  + \tilde s^2  + \tilde u^2  - 4m^2
    \tilde s + 4m^2 \tilde u) - m^2 \tilde t^2 (\tilde t - 2m^2 )} \right\}
\end{eqnarray}

 From equations (\ref{ej1}) and (\ref{ej2}), it is evident that  if $s=s_0$
and $t=t_0$ are such that they satisfy equation (\ref{deltam}) then $\tilde
 s=t_0$ and $\tilde t=s_0$ satisfy

\begin{equation}
 \left| {\tilde {\cal M} } \right|_P^2(\tilde s=t_0,\tilde t=s_0)  -
\left| {\tilde {\cal M} } \right|_A^2(\tilde s=t_0,\tilde t=s_0) =0
\end{equation}

Figures 1 and 2 show the solution of equation (\ref{deltam}) for
the systems $e^+ e^- \rightarrow \gamma \gamma$ and $W^+ W^-
\rightarrow \gamma \gamma$ respectively, in a range of $t$ and $s$
values. In both figures the continuous line shows the zeros of
equation (\ref{deltam}) that are physical for the direct process
but unphysical for crossed system ($\tilde s = t$ would be
negative). On the other hand, the pointed curve shows the zeros of
equation (\ref{deltam}) which are unphysical for the direct
process but are perfectly meaningful for the crossed one.

\section{Conclusions}

In this paper we have shown that the classical polarized
photoabsorption sum rules for the process $\gamma a \to b c$,
given in equation \ref{E3}, can be easily generalize using
detailed balance to polarized photoabsorption processes, $b c \to
\gamma a$. This means that there are values of the Mandelstam
variable $s$ for which the total cross section for the processes
where the initial photon and particle $a$ are polarized parallel
or antiparallel to each other are equal, and furthermore that
there is a range of values of the Mandelstam variables $s=s_0$ and
$t=t_0$ for which the differential cross sections for these
processes are also equal. Moreover, using crossing symmetry, we
also showed that there is a range of values of $s$ and $t$ for
which the differential cross section for the process $\gamma b \to
a c$ (or $a c \to \gamma b$) in which the helicities of the photon
and particle $a$ have specific values, is equal to the
differential cross section for the process in which one of these
two helicities is reversed.

These results could have interesting phenomenological
consequences. Perhaps a practical application is $q \bar{q} \to
\gamma^* g$ in high transverse Drell-Yan processes. Here the
photon polarization can be in principle measured via the lepton
pair angular distribution. The parallel-antiparallel spin
correlation could be that of the photon with the target proton
polarization. Another possibility is in strange quark production,
where its polarization is reflected in a leading $\Lambda$. One
example is semiinclusive DIS, where the incident photon
polarization can be controlled by the electron polarization. Of
course in these two applications we are really considering virtual
photons, and the idea is to consider the behavior of the spin
correlation as a function of photon virtuality.

\vskip 3mm
\noindent {\large{\bf{Acknowledgements}}}
\vskip 3mm

We thank Stanley Brodsky for helpful comments. This work was
supported by Fondecyt (Chile) grants No. 8000017 and 3020002
(A.Z.).

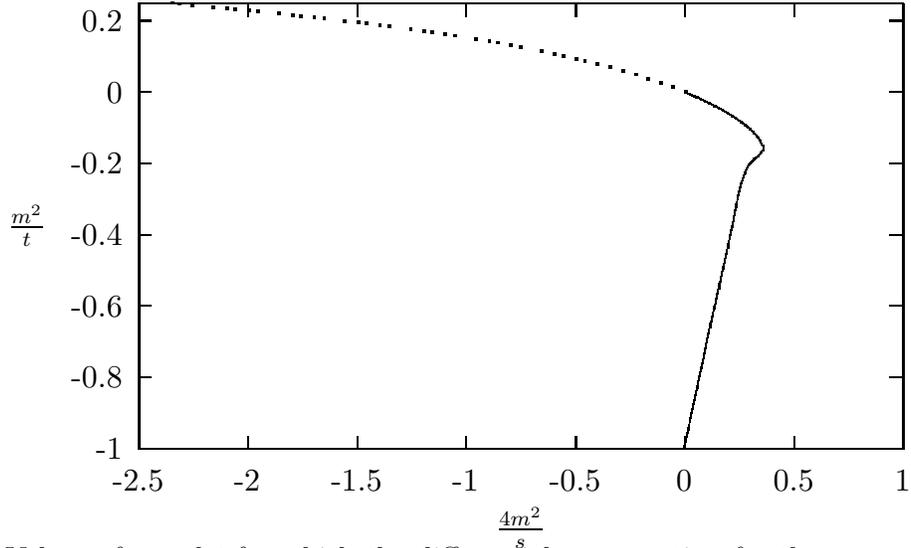
\begin{figure}[htbp]
  \begin{center}
\setlength{\unitlength}{0.240900pt}
\ifx\plotpoint\undefined\newsavebox{\plotpoint}\fi
\sbox{\plotpoint}{\rule[-0.200pt]{0.400pt}{0.400pt}}%
\begin{picture}(1500,900)(0,0)
\font\gnuplot=cmr10 at 12pt
\gnuplot
\sbox{\plotpoint}{\rule[-0.200pt]{0.400pt}{0.400pt}}%
\put(225.0,150.0){\rule[-0.200pt]{4.818pt}{0.400pt}}
\put(200,150){\makebox(0,0)[r]{-1}}
\put(1405.0,150.0){\rule[-0.200pt]{4.818pt}{0.400pt}}
\put(225.0,262.0){\rule[-0.200pt]{4.818pt}{0.400pt}}
\put(200,262){\makebox(0,0)[r]{-0.8}}
\put(1405.0,262.0){\rule[-0.200pt]{4.818pt}{0.400pt}}
\put(225.0,374.0){\rule[-0.200pt]{4.818pt}{0.400pt}}
\put(200,374){\makebox(0,0)[r]{-0.6}}
\put(1405.0,374.0){\rule[-0.200pt]{4.818pt}{0.400pt}}
\put(225.0,486.0){\rule[-0.200pt]{4.818pt}{0.400pt}}
\put(200,486){\makebox(0,0)[r]{-0.4}}
\put(1405.0,486.0){\rule[-0.200pt]{4.818pt}{0.400pt}}
\put(225.0,598.0){\rule[-0.200pt]{4.818pt}{0.400pt}}
\put(200,598){\makebox(0,0)[r]{-0.2}}
\put(1405.0,598.0){\rule[-0.200pt]{4.818pt}{0.400pt}}
\put(225.0,710.0){\rule[-0.200pt]{4.818pt}{0.400pt}}
\put(200,710){\makebox(0,0)[r]{0}}
\put(1405.0,710.0){\rule[-0.200pt]{4.818pt}{0.400pt}}
\put(225.0,822.0){\rule[-0.200pt]{4.818pt}{0.400pt}}
\put(200,822){\makebox(0,0)[r]{0.2}}
\put(1405.0,822.0){\rule[-0.200pt]{4.818pt}{0.400pt}}
\put(225.0,150.0){\rule[-0.200pt]{0.400pt}{4.818pt}}
\put(225,100){\makebox(0,0){-2.5}}
\put(225.0,830.0){\rule[-0.200pt]{0.400pt}{4.818pt}}
\put(396.0,150.0){\rule[-0.200pt]{0.400pt}{4.818pt}}
\put(396,100){\makebox(0,0){-2}}
\put(396.0,830.0){\rule[-0.200pt]{0.400pt}{4.818pt}}
\put(568.0,150.0){\rule[-0.200pt]{0.400pt}{4.818pt}}
\put(568,100){\makebox(0,0){-1.5}}
\put(568.0,830.0){\rule[-0.200pt]{0.400pt}{4.818pt}}
\put(739.0,150.0){\rule[-0.200pt]{0.400pt}{4.818pt}}
\put(739,100){\makebox(0,0){-1}}
\put(739.0,830.0){\rule[-0.200pt]{0.400pt}{4.818pt}}
\put(911.0,150.0){\rule[-0.200pt]{0.400pt}{4.818pt}}
\put(911,100){\makebox(0,0){-0.5}}
\put(911.0,830.0){\rule[-0.200pt]{0.400pt}{4.818pt}}
\put(1082.0,150.0){\rule[-0.200pt]{0.400pt}{4.818pt}}
\put(1082,100){\makebox(0,0){0}}
\put(1082.0,830.0){\rule[-0.200pt]{0.400pt}{4.818pt}}
\put(1254.0,150.0){\rule[-0.200pt]{0.400pt}{4.818pt}}
\put(1254,100){\makebox(0,0){0.5}}
\put(1254.0,830.0){\rule[-0.200pt]{0.400pt}{4.818pt}}
\put(1425.0,150.0){\rule[-0.200pt]{0.400pt}{4.818pt}}
\put(1425,100){\makebox(0,0){1}}
\put(1425.0,830.0){\rule[-0.200pt]{0.400pt}{4.818pt}}
\put(225.0,150.0){\rule[-0.200pt]{289.080pt}{0.400pt}}
\put(1425.0,150.0){\rule[-0.200pt]{0.400pt}{168.630pt}}
\put(225.0,850.0){\rule[-0.200pt]{289.080pt}{0.400pt}}
\put(50,500){\makebox(0,0){$\frac{m^2}{t}$}}
\put(825,25){\makebox(0,0){$\frac{4m^2}{s}$}}
\put(225.0,150.0){\rule[-0.200pt]{0.400pt}{168.630pt}}
\put(1082,150){\usebox{\plotpoint}}
\multiput(1082.59,150.00)(0.485,2.476){11}{\rule{0.117pt}{1.986pt}}
\multiput(1081.17,150.00)(7.000,28.879){2}{\rule{0.400pt}{0.993pt}}
\multiput(1089.59,183.00)(0.485,2.476){11}{\rule{0.117pt}{1.986pt}}
\multiput(1088.17,183.00)(7.000,28.879){2}{\rule{0.400pt}{0.993pt}}
\multiput(1096.59,216.00)(0.485,2.476){11}{\rule{0.117pt}{1.986pt}}
\multiput(1095.17,216.00)(7.000,28.879){2}{\rule{0.400pt}{0.993pt}}
\multiput(1103.59,249.00)(0.485,2.323){11}{\rule{0.117pt}{1.871pt}}
\multiput(1102.17,249.00)(7.000,27.116){2}{\rule{0.400pt}{0.936pt}}
\multiput(1110.59,280.00)(0.482,2.751){9}{\rule{0.116pt}{2.167pt}}
\multiput(1109.17,280.00)(6.000,26.503){2}{\rule{0.400pt}{1.083pt}}
\multiput(1116.59,311.00)(0.485,2.399){11}{\rule{0.117pt}{1.929pt}}
\multiput(1115.17,311.00)(7.000,27.997){2}{\rule{0.400pt}{0.964pt}}
\multiput(1123.59,343.00)(0.485,2.247){11}{\rule{0.117pt}{1.814pt}}
\multiput(1122.17,343.00)(7.000,26.234){2}{\rule{0.400pt}{0.907pt}}
\multiput(1130.59,373.00)(0.485,2.399){11}{\rule{0.117pt}{1.929pt}}
\multiput(1129.17,373.00)(7.000,27.997){2}{\rule{0.400pt}{0.964pt}}
\multiput(1137.59,405.00)(0.485,2.323){11}{\rule{0.117pt}{1.871pt}}
\multiput(1136.17,405.00)(7.000,27.116){2}{\rule{0.400pt}{0.936pt}}
\multiput(1144.59,436.00)(0.485,2.399){11}{\rule{0.117pt}{1.929pt}}
\multiput(1143.17,436.00)(7.000,27.997){2}{\rule{0.400pt}{0.964pt}}
\multiput(1151.59,468.00)(0.485,2.476){11}{\rule{0.117pt}{1.986pt}}
\multiput(1150.17,468.00)(7.000,28.879){2}{\rule{0.400pt}{0.993pt}}
\multiput(1158.59,501.00)(0.482,2.932){9}{\rule{0.116pt}{2.300pt}}
\multiput(1157.17,501.00)(6.000,28.226){2}{\rule{0.400pt}{1.150pt}}
\multiput(1164.59,534.00)(0.485,2.247){11}{\rule{0.117pt}{1.814pt}}
\multiput(1163.17,534.00)(7.000,26.234){2}{\rule{0.400pt}{0.907pt}}
\multiput(1171.59,564.00)(0.485,1.560){11}{\rule{0.117pt}{1.300pt}}
\multiput(1170.17,564.00)(7.000,18.302){2}{\rule{0.400pt}{0.650pt}}
\multiput(1178.59,585.00)(0.485,0.950){11}{\rule{0.117pt}{0.843pt}}
\multiput(1177.17,585.00)(7.000,11.251){2}{\rule{0.400pt}{0.421pt}}
\multiput(1185.00,598.59)(0.492,0.485){11}{\rule{0.500pt}{0.117pt}}
\multiput(1185.00,597.17)(5.962,7.000){2}{\rule{0.250pt}{0.400pt}}
\multiput(1192.00,605.59)(0.710,0.477){7}{\rule{0.660pt}{0.115pt}}
\multiput(1192.00,604.17)(5.630,5.000){2}{\rule{0.330pt}{0.400pt}}
\multiput(1199.59,610.00)(0.485,0.645){11}{\rule{0.117pt}{0.614pt}}
\multiput(1198.17,610.00)(7.000,7.725){2}{\rule{0.400pt}{0.307pt}}
\put(1206.0,619.0){\rule[-0.200pt]{0.400pt}{1.445pt}}
\put(1082,710){\usebox{\plotpoint}}
\multiput(1082.00,708.95)(1.355,-0.447){3}{\rule{1.033pt}{0.108pt}}
\multiput(1082.00,709.17)(4.855,-3.000){2}{\rule{0.517pt}{0.400pt}}
\multiput(1089.00,705.95)(1.355,-0.447){3}{\rule{1.033pt}{0.108pt}}
\multiput(1089.00,706.17)(4.855,-3.000){2}{\rule{0.517pt}{0.400pt}}
\put(1096,702.17){\rule{1.500pt}{0.400pt}}
\multiput(1096.00,703.17)(3.887,-2.000){2}{\rule{0.750pt}{0.400pt}}
\multiput(1103.00,700.94)(0.920,-0.468){5}{\rule{0.800pt}{0.113pt}}
\multiput(1103.00,701.17)(5.340,-4.000){2}{\rule{0.400pt}{0.400pt}}
\multiput(1110.00,696.95)(1.132,-0.447){3}{\rule{0.900pt}{0.108pt}}
\multiput(1110.00,697.17)(4.132,-3.000){2}{\rule{0.450pt}{0.400pt}}
\multiput(1116.00,693.95)(1.355,-0.447){3}{\rule{1.033pt}{0.108pt}}
\multiput(1116.00,694.17)(4.855,-3.000){2}{\rule{0.517pt}{0.400pt}}
\multiput(1123.00,690.94)(0.920,-0.468){5}{\rule{0.800pt}{0.113pt}}
\multiput(1123.00,691.17)(5.340,-4.000){2}{\rule{0.400pt}{0.400pt}}
\multiput(1130.00,686.94)(0.920,-0.468){5}{\rule{0.800pt}{0.113pt}}
\multiput(1130.00,687.17)(5.340,-4.000){2}{\rule{0.400pt}{0.400pt}}
\multiput(1137.00,682.95)(1.355,-0.447){3}{\rule{1.033pt}{0.108pt}}
\multiput(1137.00,683.17)(4.855,-3.000){2}{\rule{0.517pt}{0.400pt}}
\multiput(1144.00,679.93)(0.710,-0.477){7}{\rule{0.660pt}{0.115pt}}
\multiput(1144.00,680.17)(5.630,-5.000){2}{\rule{0.330pt}{0.400pt}}
\multiput(1151.00,674.94)(0.920,-0.468){5}{\rule{0.800pt}{0.113pt}}
\multiput(1151.00,675.17)(5.340,-4.000){2}{\rule{0.400pt}{0.400pt}}
\multiput(1158.00,670.94)(0.774,-0.468){5}{\rule{0.700pt}{0.113pt}}
\multiput(1158.00,671.17)(4.547,-4.000){2}{\rule{0.350pt}{0.400pt}}
\multiput(1164.00,666.93)(0.710,-0.477){7}{\rule{0.660pt}{0.115pt}}
\multiput(1164.00,667.17)(5.630,-5.000){2}{\rule{0.330pt}{0.400pt}}
\multiput(1171.00,661.93)(0.710,-0.477){7}{\rule{0.660pt}{0.115pt}}
\multiput(1171.00,662.17)(5.630,-5.000){2}{\rule{0.330pt}{0.400pt}}
\multiput(1178.00,656.93)(0.581,-0.482){9}{\rule{0.567pt}{0.116pt}}
\multiput(1178.00,657.17)(5.824,-6.000){2}{\rule{0.283pt}{0.400pt}}
\multiput(1185.00,650.93)(0.581,-0.482){9}{\rule{0.567pt}{0.116pt}}
\multiput(1185.00,651.17)(5.824,-6.000){2}{\rule{0.283pt}{0.400pt}}
\multiput(1192.59,643.69)(0.485,-0.569){11}{\rule{0.117pt}{0.557pt}}
\multiput(1191.17,644.84)(7.000,-6.844){2}{\rule{0.400pt}{0.279pt}}
\multiput(1199.59,634.50)(0.485,-0.950){11}{\rule{0.117pt}{0.843pt}}
\multiput(1198.17,636.25)(7.000,-11.251){2}{\rule{0.400pt}{0.421pt}}
\sbox{\plotpoint}{\rule[-0.500pt]{1.000pt}{1.000pt}}%
\put(1082,710){\usebox{\plotpoint}}
\put(1082.00,710.00){\usebox{\plotpoint}}
\put(1062.76,717.75){\usebox{\plotpoint}}
\put(1043.14,724.39){\usebox{\plotpoint}}
\put(1023.50,731.00){\usebox{\plotpoint}}
\put(1003.99,737.86){\usebox{\plotpoint}}
\put(983.93,742.89){\usebox{\plotpoint}}
\put(964.25,749.37){\usebox{\plotpoint}}
\put(944.09,754.26){\usebox{\plotpoint}}
\put(923.90,759.02){\usebox{\plotpoint}}
\multiput(911,762)(-20.234,4.625){2}{\usebox{\plotpoint}}
\multiput(876,770)(-20.329,4.185){2}{\usebox{\plotpoint}}
\put(822.49,781.02){\usebox{\plotpoint}}
\multiput(808,784)(-20.440,3.607){2}{\usebox{\plotpoint}}
\multiput(774,790)(-20.457,3.507){2}{\usebox{\plotpoint}}
\put(720.25,798.76){\usebox{\plotpoint}}
\multiput(705,801)(-20.535,3.020){2}{\usebox{\plotpoint}}
\multiput(671,806)(-20.547,2.935){2}{\usebox{\plotpoint}}
\put(617.49,813.18){\usebox{\plotpoint}}
\multiput(602,815)(-20.535,3.020){2}{\usebox{\plotpoint}}
\multiput(568,820)(-20.675,1.824){2}{\usebox{\plotpoint}}
\put(514.51,825.78){\usebox{\plotpoint}}
\multiput(499,828)(-20.675,1.824){2}{\usebox{\plotpoint}}
\multiput(465,831)(-20.613,2.425){2}{\usebox{\plotpoint}}
\put(411.33,836.69){\usebox{\plotpoint}}
\multiput(396,838)(-20.613,2.425){2}{\usebox{\plotpoint}}
\multiput(362,842)(-20.675,1.824){2}{\usebox{\plotpoint}}
\put(308.05,846.76){\usebox{\plotpoint}}
\put(287.39,848.73){\usebox{\plotpoint}}
\put(276,850){\usebox{\plotpoint}}
\end{picture}
    \caption{Values of $s$ and $t$ for which the differential
cross section for the process in which the helicities of the
photon and particle $a$ have specific values is equal to the
differential cross section for the process in which one of the two
helicities is reversed. In the particular case shown here $a$ is
also a photon. So the continuous line shows the zero for $e^+ e^-
\to \gamma \gamma$, and the dotted line for the crossed process
$\gamma e^- \to \gamma e^-$ ($m$ is the electron mass).}
    \label{fig:1}
  \end{center}
\end{figure}

\begin{figure}[htbp]
  \begin{center}
\setlength{\unitlength}{0.240900pt}
\ifx\plotpoint\undefined\newsavebox{\plotpoint}\fi
\sbox{\plotpoint}{\rule[-0.200pt]{0.400pt}{0.400pt}}%
\begin{picture}(1500,900)(0,0)
\font\gnuplot=cmr10 at 12pt
\gnuplot
\sbox{\plotpoint}{\rule[-0.200pt]{0.400pt}{0.400pt}}%
\put(225.0,150.0){\rule[-0.200pt]{4.818pt}{0.400pt}}
\put(200,150){\makebox(0,0)[r]{-1}}
\put(1405.0,150.0){\rule[-0.200pt]{4.818pt}{0.400pt}}
\put(225.0,250.0){\rule[-0.200pt]{4.818pt}{0.400pt}}
\put(200,250){\makebox(0,0)[r]{-0.8}}
\put(1405.0,250.0){\rule[-0.200pt]{4.818pt}{0.400pt}}
\put(225.0,350.0){\rule[-0.200pt]{4.818pt}{0.400pt}}
\put(200,350){\makebox(0,0)[r]{-0.6}}
\put(1405.0,350.0){\rule[-0.200pt]{4.818pt}{0.400pt}}
\put(225.0,450.0){\rule[-0.200pt]{4.818pt}{0.400pt}}
\put(200,450){\makebox(0,0)[r]{-0.4}}
\put(1405.0,450.0){\rule[-0.200pt]{4.818pt}{0.400pt}}
\put(225.0,550.0){\rule[-0.200pt]{4.818pt}{0.400pt}}
\put(200,550){\makebox(0,0)[r]{-0.2}}
\put(1405.0,550.0){\rule[-0.200pt]{4.818pt}{0.400pt}}
\put(225.0,650.0){\rule[-0.200pt]{4.818pt}{0.400pt}}
\put(200,650){\makebox(0,0)[r]{0}}
\put(1405.0,650.0){\rule[-0.200pt]{4.818pt}{0.400pt}}
\put(225.0,750.0){\rule[-0.200pt]{4.818pt}{0.400pt}}
\put(200,750){\makebox(0,0)[r]{0.2}}
\put(1405.0,750.0){\rule[-0.200pt]{4.818pt}{0.400pt}}
\put(225.0,850.0){\rule[-0.200pt]{4.818pt}{0.400pt}}
\put(200,850){\makebox(0,0)[r]{0.4}}
\put(1405.0,850.0){\rule[-0.200pt]{4.818pt}{0.400pt}}
\put(225.0,150.0){\rule[-0.200pt]{0.400pt}{4.818pt}}
\put(225,100){\makebox(0,0){-2.5}}
\put(225.0,830.0){\rule[-0.200pt]{0.400pt}{4.818pt}}
\put(396.0,150.0){\rule[-0.200pt]{0.400pt}{4.818pt}}
\put(396,100){\makebox(0,0){-2}}
\put(396.0,830.0){\rule[-0.200pt]{0.400pt}{4.818pt}}
\put(568.0,150.0){\rule[-0.200pt]{0.400pt}{4.818pt}}
\put(568,100){\makebox(0,0){-1.5}}
\put(568.0,830.0){\rule[-0.200pt]{0.400pt}{4.818pt}}
\put(739.0,150.0){\rule[-0.200pt]{0.400pt}{4.818pt}}
\put(739,100){\makebox(0,0){-1}}
\put(739.0,830.0){\rule[-0.200pt]{0.400pt}{4.818pt}}
\put(911.0,150.0){\rule[-0.200pt]{0.400pt}{4.818pt}}
\put(911,100){\makebox(0,0){-0.5}}
\put(911.0,830.0){\rule[-0.200pt]{0.400pt}{4.818pt}}
\put(1082.0,150.0){\rule[-0.200pt]{0.400pt}{4.818pt}}
\put(1082,100){\makebox(0,0){0}}
\put(1082.0,830.0){\rule[-0.200pt]{0.400pt}{4.818pt}}
\put(1254.0,150.0){\rule[-0.200pt]{0.400pt}{4.818pt}}
\put(1254,100){\makebox(0,0){0.5}}
\put(1254.0,830.0){\rule[-0.200pt]{0.400pt}{4.818pt}}
\put(1425.0,150.0){\rule[-0.200pt]{0.400pt}{4.818pt}}
\put(1425,100){\makebox(0,0){1}}
\put(1425.0,830.0){\rule[-0.200pt]{0.400pt}{4.818pt}}
\put(225.0,150.0){\rule[-0.200pt]{289.080pt}{0.400pt}}
\put(1425.0,150.0){\rule[-0.200pt]{0.400pt}{168.630pt}}
\put(225.0,850.0){\rule[-0.200pt]{289.080pt}{0.400pt}}
\put(50,500){\makebox(0,0){$\frac{M^2}{t}$}}
\put(825,25){\makebox(0,0){$\frac{4M^2}{s}$}}
\put(225.0,150.0){\rule[-0.200pt]{0.400pt}{168.630pt}}
\put(1082,150){\usebox{\plotpoint}}
\multiput(1082.58,150.00)(0.498,0.737){65}{\rule{0.120pt}{0.688pt}}
\multiput(1081.17,150.00)(34.000,48.572){2}{\rule{0.400pt}{0.344pt}}
\multiput(1116.58,200.00)(0.498,0.715){67}{\rule{0.120pt}{0.671pt}}
\multiput(1115.17,200.00)(35.000,48.606){2}{\rule{0.400pt}{0.336pt}}
\multiput(1151.58,250.00)(0.498,0.811){65}{\rule{0.120pt}{0.747pt}}
\multiput(1150.17,250.00)(34.000,53.449){2}{\rule{0.400pt}{0.374pt}}
\multiput(1185.58,305.00)(0.498,0.885){65}{\rule{0.120pt}{0.806pt}}
\multiput(1184.17,305.00)(34.000,58.327){2}{\rule{0.400pt}{0.403pt}}
\multiput(1219.59,365.00)(0.485,1.484){11}{\rule{0.117pt}{1.243pt}}
\multiput(1218.17,365.00)(7.000,17.420){2}{\rule{0.400pt}{0.621pt}}
\multiput(1226.58,385.00)(0.491,1.017){17}{\rule{0.118pt}{0.900pt}}
\multiput(1225.17,385.00)(10.000,18.132){2}{\rule{0.400pt}{0.450pt}}
\multiput(1236.59,405.00)(0.485,1.484){11}{\rule{0.117pt}{1.243pt}}
\multiput(1235.17,405.00)(7.000,17.420){2}{\rule{0.400pt}{0.621pt}}
\multiput(1243.60,425.00)(0.468,1.358){5}{\rule{0.113pt}{1.100pt}}
\multiput(1242.17,425.00)(4.000,7.717){2}{\rule{0.400pt}{0.550pt}}
\multiput(1247.61,435.00)(0.447,3.141){3}{\rule{0.108pt}{2.100pt}}
\multiput(1246.17,435.00)(3.000,10.641){2}{\rule{0.400pt}{1.050pt}}
\multiput(1250.60,450.00)(0.468,5.014){5}{\rule{0.113pt}{3.600pt}}
\multiput(1249.17,450.00)(4.000,27.528){2}{\rule{0.400pt}{1.800pt}}
\put(1082,650){\usebox{\plotpoint}}
\multiput(1082.00,648.92)(1.329,-0.493){23}{\rule{1.146pt}{0.119pt}}
\multiput(1082.00,649.17)(31.621,-13.000){2}{\rule{0.573pt}{0.400pt}}
\multiput(1116.00,635.92)(1.039,-0.495){31}{\rule{0.924pt}{0.119pt}}
\multiput(1116.00,636.17)(33.083,-17.000){2}{\rule{0.462pt}{0.400pt}}
\multiput(1151.00,618.92)(0.855,-0.496){37}{\rule{0.780pt}{0.119pt}}
\multiput(1151.00,619.17)(32.381,-20.000){2}{\rule{0.390pt}{0.400pt}}
\multiput(1185.00,598.92)(0.566,-0.497){57}{\rule{0.553pt}{0.120pt}}
\multiput(1185.00,599.17)(32.852,-30.000){2}{\rule{0.277pt}{0.400pt}}
\multiput(1219.59,567.21)(0.485,-0.721){11}{\rule{0.117pt}{0.671pt}}
\multiput(1218.17,568.61)(7.000,-8.606){2}{\rule{0.400pt}{0.336pt}}
\multiput(1226.00,558.92)(0.495,-0.491){17}{\rule{0.500pt}{0.118pt}}
\multiput(1226.00,559.17)(8.962,-10.000){2}{\rule{0.250pt}{0.400pt}}
\multiput(1236.59,546.03)(0.485,-1.103){11}{\rule{0.117pt}{0.957pt}}
\multiput(1235.17,548.01)(7.000,-13.013){2}{\rule{0.400pt}{0.479pt}}
\multiput(1243.60,530.43)(0.468,-1.358){5}{\rule{0.113pt}{1.100pt}}
\multiput(1242.17,532.72)(4.000,-7.717){2}{\rule{0.400pt}{0.550pt}}
\multiput(1247.61,519.05)(0.447,-2.025){3}{\rule{0.108pt}{1.433pt}}
\multiput(1246.17,522.03)(3.000,-7.025){2}{\rule{0.400pt}{0.717pt}}
\multiput(1250.60,502.13)(0.468,-4.283){5}{\rule{0.113pt}{3.100pt}}
\multiput(1249.17,508.57)(4.000,-23.566){2}{\rule{0.400pt}{1.550pt}}
\sbox{\plotpoint}{\rule[-0.500pt]{1.000pt}{1.000pt}}%
\put(1082,650){\usebox{\plotpoint}}
\multiput(1082,650)(-19.912,5.857){4}{\usebox{\plotpoint}}
\multiput(1014,670)(-19.935,5.778){3}{\usebox{\plotpoint}}
\multiput(945,690)(-20.541,2.977){4}{\usebox{\plotpoint}}
\multiput(876,700)(-20.268,4.471){3}{\usebox{\plotpoint}}
\multiput(808,715)(-20.541,2.977){3}{\usebox{\plotpoint}}
\multiput(739,725)(-20.535,3.020){4}{\usebox{\plotpoint}}
\multiput(671,735)(-20.541,2.977){3}{\usebox{\plotpoint}}
\multiput(602,745)(-20.700,1.522){3}{\usebox{\plotpoint}}
\multiput(534,750)(-20.541,2.977){4}{\usebox{\plotpoint}}
\multiput(465,760)(-20.541,2.977){3}{\usebox{\plotpoint}}
\multiput(396,770)(-20.700,1.522){3}{\usebox{\plotpoint}}
\multiput(328,775)(-20.701,1.500){4}{\usebox{\plotpoint}}
\put(243.32,781.38){\usebox{\plotpoint}}
\put(225,783){\usebox{\plotpoint}}
\end{picture}

    \caption{Same as figure 1, but for the process $W^+ W^- \to \gamma \gamma$
($M$ is the $W$ mass).}
    \label{fig:2}
  \end{center}
\end{figure}
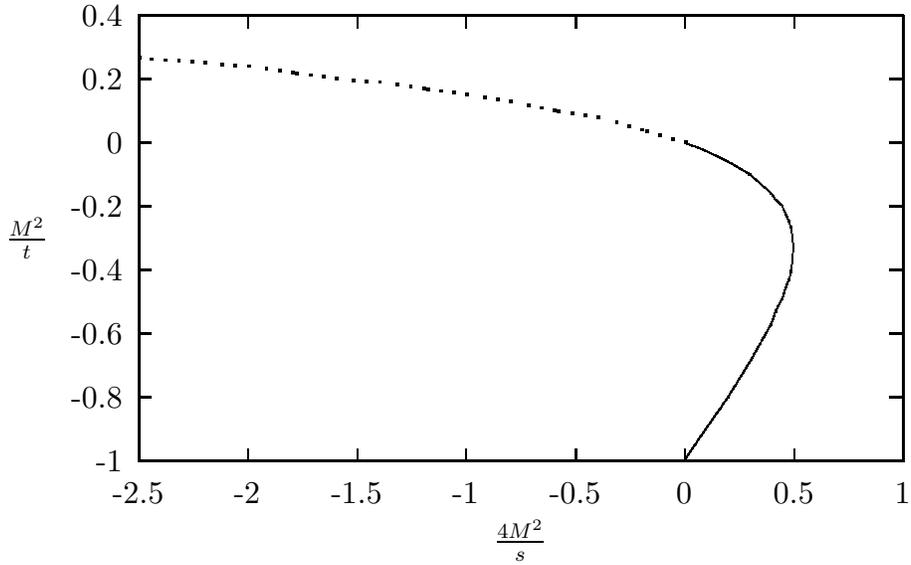


\begin{thebibliography}{99}

\bibitem{LlewellynSmith:1998wc}
C.~H.~Llewellyn Smith,
arXiv:hep-ph/9812301.

\bibitem{dh}D. Drell and A.C. Hearn, \PRL{16}{66}{908}.

\bibitem{gerasimov} S. Gerasimov, {\it Sov. J. Nucl. Phys.} {\bf 2}(1966)430.

\bibitem{altarelli}G. Altarelli, N. Cabibbo and L. Maiani, \PLB{40}{72}{415}.

\bibitem{bs}S. J. Brodsky and I. Schmidt, \PLB{351}{95}{344}.


\bibitem{Brodsky:1995ga}
S.~J.~Brodsky, T.~G.~Rizzo and I.~Schmidt,
Phys.\ Rev.\ D {\bf 52}, 4929 (1995) [arXiv:hep-ph/9505441].

\bibitem{Brodsky:1997ex}
S.~J.~Brodsky and I.~Schmidt,
Phys.\ Lett.\ B {\bf 423}, 145 (1998) [arXiv:hep-ph/9711413].

\bibitem{Bass:1998rn}
S.~D.~Bass, S.~J.~Brodsky and I.~Schmidt,
Phys.\ Rev.\ D {\bf 60}, 034010 (1999) [arXiv:hep-ph/9901244].

\bibitem{Bass:1998bw}
S.~D.~Bass, S.~J.~Brodsky and I.~Schmidt,
Phys.\ Lett.\ B {\bf 437}, 417 (1998) [arXiv:hep-ph/9805316].

\bibitem{Goldberg:1999gc}
H.~Goldberg,
Phys.\ Lett.\ B {\bf 472}, 280 (2000) [arXiv:hep-ph/9904318].

\bibitem{martin}A.D. Martin and T.D. Spearman, ``Elementary Particle
  Theory'' (North-Holland Publishing Company, Amsterdam, 1970)
\end{thebibliography}
\end{document}